\documentclass
[floatfix,superscriptaddress,secnumarabic,amssymb,amsmath,nobibnotes,aps,prd,showkeys,showpacs,nofootinbib,onecolumn,notitlepage,12pt]{revtex4}%
\usepackage{setspace}
\usepackage{color}
\usepackage{amsmath}
\usepackage{amsfonts}
\usepackage{verbatim}
\usepackage{amssymb}
\usepackage{graphicx,bm}
\usepackage{graphicx}
\usepackage{amsmath}
\usepackage{amssymb}
\usepackage{amssymb}
\usepackage{graphicx,bm}
\usepackage{graphicx}
\usepackage[caption=false]{subfig}%
\providecommand{\U}[1]{\protect\rule{.1in}{.1in}}

\newcommand{\be}{\begin{equation}}
\newcommand{\ee}{\end{equation}}

\newcommand{\mincir}{\raise
-3.truept\hbox{\rlap{\hbox{$\sim$}}\raise4.truept\hbox{$<$}\ }}
\newcommand{\magcir}{\raise
-3.truept\hbox{\rlap{\hbox{$\sim$}}\raise4.truept\hbox{$>$}\ }}

\ifx\pdfoutput\relax\let\pdfoutput=\undefined\fi
\newcount\msipdfoutput
\ifx\pdfoutput\undefined\else
\ifcase\pdfoutput\else
\ifx\paperwidth\undefined\else
\ifdim\paperheight=0pt\relax\else\pdfpageheight\paperheight\fi
\ifdim\paperwidth=0pt\relax\else\pdfpagewidth\paperwidth\fi
\fi\fi\fi
\begin{document}
\title{The Brans-Dicke field in Non-metricity gravity: Cosmological solutions and
Conformal transformations}
\author{Andronikos Paliathanasis}
\email{anpaliat@phys.uoa.gr}
\affiliation{Institute of Systems Science, Durban University of Technology, Durban 4000,
South Africa}
\affiliation{Departamento de Matem\'{a}ticas, Universidad Cat\'{o}lica del Norte, Avda.
Angamos 0610, Casilla 1280 Antofagasta, Chile}

\begin{abstract}
We consider the Brans-Dicke theory in non-metricity gravity, which belongs to
the family of symmetric teleparallel scalar-tensor theories. Our focus lies in
exploring the implications of the conformal transformation, as we derive the
conformal equivalent theory in the Einstein frame, distinct from the minimally
coupled scalar field theory. The fundamental principle of the conformal
transformation suggests the mathematical equivalence of the related theories.
However, to thoroughly analyze the impact on physical variables, we
investigate the spatially flat Friedmann--Lema\^{\i}tre--Robertson--Walker
geometry, defining the connection in the non-coincidence gauge. We construct
exact solutions for the cosmological model in one frame and compare the
physical properties in the conformal related frame. Surprisingly, we find that
the general physical properties of the exact solutions remain invariant under
the conformal transformation. Finally, we construct, for the first time, an
analytic solution for the symmetric teleparallel scalar-tensor cosmology.

\end{abstract}
\keywords{Brans-Dicke field; non-metricity gravity; scalar-tensor cosmology; conformal transformation.}\maketitle

\section{Introduction}

\label{sec1}

Symmetric Teleparallel General Relativity (STGR) \cite{Nester:1998mp}
represents an alternative gravitational theory, considered equivalent to
General Relativity (GR). In STGR, the fundamental geometric elements consist
of the metric tensor $g_{\mu\nu}$ and the symmetric, flat connection
$\Gamma_{\mu\nu}^{\lambda}$ with the covariant derivative $\nabla_{\lambda}$,
leading to $\nabla_{\lambda}g_{\mu\nu}\neq0$. While GR defines autoparallels
using the Levi-Civita connection for the metric tensor $g_{\mu\nu}$, STGR
emphasizes the non-metricity component, crucial for the theory's description.
The equivalence between these two gravitational theories becomes evident upon
a study of the gravitational Lagrangians \cite{lav2}. In GR, the Lagrangian
function involves the Ricci scalar constructed by the Levi-Civita connection
$\mathring{R}$, whereas in STGR, the corresponding Lagrangian is defined by
the non-metricity scalar $Q$. The Ricci scalar and the non-metricity scalar
differ by a boundary term $B=\mathring{R}-Q$ \cite{Nester:1998mp,lav2,Hohmann}%
. Consequently, the variation of the two distinct Lagrangians yields the same
physical theory. However, this equivalence breaks down when introducing matter
non-minimally coupled to gravity \cite{sc1,gg1,st1}, or nonlinear terms of the
gravitational scalars in the Action Integral \cite{Koivisto2,Koivisto3}.

In $f(Q)$-gravity \cite{Koivisto2,Koivisto3}, a straightforward extension of
the STGR theory, the gravitational Lagrangian takes the form of a nonlinear
function $f$ of the non-metricity scalar $Q$. These nonlinear terms introduce
additional degrees of freedom, leading to modifications in the gravitational
field equations that give rise to new phenomena \cite{rev10}. In the context
of cosmology, $f(Q)$ has been proposed as a solution to the dark energy
problem \cite{ft1,ft2,ft3,ft4,ft5,ft6} and has been utilized to explain cosmic
acceleration \cite{rr1,Teg,Kowal,Komatsu}.

In the symmetric teleparallel theory of gravity, the presence of a flat
geometry defined by the connection $\Gamma_{\mu\nu}^{\lambda}$ allows for the
existence of a coordinate system known as the coincidence gauge, where the
covariant derivative can be represented as a partial derivative. This implies
that in the symmetric teleparallel theory of gravity, the inertial effects can
be distinguished from gravity. Consequently, the choice of the connection as
the starting point in the symmetric teleparallel theory leads to the
formulation of distinct gravitational theories \cite{Hohmann}. As a result,
self-accelerating solutions can naturally emerge both in the early and late
universe \cite{ndim}. The impact of different connections on the existence of
cosmological solutions has been extensively explored in \cite{ndim}, while the
scenario of static spherically symmetric spacetimes has been considered in
\cite{bah1,bah2}. The reconstruction of the cosmological history was derived
in \cite{an0,an1,an2}. Specifically, the phase-space analysis was studied, for
the field equations for the four different connections which describe the
Friedmann--Lema\^{\i}tre--Robertson--Walker (FLRW) geometry \cite{Hohmann}.
For similar studies see also \cite{dn1,dn2}. Quantum cosmology in $f\left(
Q\right)  $-gravity investigated in \cite{ndim3}, while in \cite{ndim2} a
minisuperspace description is presented from where it follows that the
$f\left(  Q\right)  $-theory can be described by two scalar fields. The first
scalar field corresponds to the degrees of freedom associated with the
higher-order derivatives of the theory, whereas the second scalar field is
linked to the connection defined in the non-coincidence gauge. For further
investigations into $f(Q)$-gravity and its generalizations, we recommend
referring to the works cited in
\cite{fq1,fq2,fq3,fq4,fq5,fq6,fq7,fq8,fq9,fbo1,fq10} and the references
provided therein.

Scalar fields non-minimally coupled to gravity have found extensive
application in gravitational physics within the framework of General
Relativity, such as in scalar-curvature theories \cite{sf1,sf2}, or in the
context of teleparallelism, specifically scalar-tensor theories \cite{sf3,sf4}%
. The Brans-Dicke theory \cite{Brans} represents one of the earliest
scalar-curvature theories, formulated with the intention of establishing a
gravitational theory that adheres to Mach's principle. This model is defined
in the Jordan frame \cite{Jord}, where the presence of a matter source is
essential for the existence of physical space. In contrast, General Relativity
is defined in the Einstein frame, enabling the existence of physical space
even in the absence of a matter term. The Brans-Dicke parameter is a
characterized constant of the theory which indicates the coupling between the
scalar and the gravitation Lagrangian \cite{omegaBDGR}. When the Bans-Dicke
parameter vanishes, the theory is equivalent to the $f\left(  R\right)
$-gravity, where the non-minimally coupled scalar field attributes the
higher-order degrees of freedom \cite{s01}.

Although the scalar-curvature theory is initially defined in the Jordan frame,
a geometrical mapping exists that enables the transformation of the theory
into the Einstein frame. Consequently, the scalar-curvature theory can be
interpreted in the form equivalent to General Relativity, involving a
minimally coupled scalar field. This geometric mapping is a conformal
transformation, establishing a connection between the solution trajectories of
the two frames \cite{bb00}. However, the physical properties of the solution
trajectories are not invariant under the application of the conformal
transformation. For example singular solutions does not remain singulars after
the application of the conformal transformation, for more details see the
discussion \cite{bb02,bb03,bb04} and references therein. More recently, the
Hamiltonian inequivalence between the Jordan and Einstein frames has been
explored in \cite{gio1,gio2,gio3}.

In this study we are interested to study the effects of the conformal
transformation on the physical properties of cosmological solutions on the
Brans-Dicke analogue in symmetric teleparallel scalar-tensor theory
\cite{sc1}. It is known that $f\left(  Q\right)  $-gravity is equivalent to a
specific family of symmetric teleparallel scalar-tensor models, and we use the
analogy of the Brans-Dicke model with the $f\left(  R\right)  $-gravity in
order to introduce the non-metricity Brans-Dicke theory. We focus\ in the
cosmological scenario of a spatially flat FLRW geometry. Moreover, we consider
the case in which the connection is defined in the non-coincidence gauge and
the gravitational theory is equivalent to a multiscalar field model. While the
mathematical application of the conformal transformation in non-metricity
theory has been previously explored in \cite{gg1}, no concrete conclusions
were drawn regarding the physical properties of the solutions under the
conformal transformation. More recently, in \cite{pat1}, several exact
cosmological solutions were identified in the non-metricity scalar-tensor
theory for the non-coincidence gauge. Within this work, we aim to determine
exact and analytic solutions for the non-metricity Brans-Dicke cosmological
theory, subsequently comparing the physical properties of the solutions
between the Jordan and the Einstein frames. The structure of the paper is
outlined as follows.

In Section \ref{sec2} we discuss the fundamental properties and definitions of
symmetric teleparallel gravity. Additionally, we explore $f\left(  Q\right)
$-theory and the symmetric teleparallel scalar-tensor theory of gravity. We
demonstrate that $f\left(  Q\right)  $-theory can be reformulated as a
non-metricity scalar-tensor theory. Furthermore, we present the utilization of
conformal transformations and the derivation of the conformal equivalent
theory in Section \ref{sec2a}. In Section \ref{sec3}, we introduce the
extension of the Brans-Dicke field in non-metricity gravity. Here, we
introduce a novel parameter $\omega$, akin to the Brans-Dicke parameter of
scalar-curvature theory. As $\omega\rightarrow0$, the gravitational Action
characterizes the $f\left(  Q\right)  $-theory, similarly to how the
Brans-Dicke field characterizes $f\left(  R\right)  $-gravity in the same
limit. Within this gravitational model, we consider a spatially flat FLRW
background geometry, and for the connection defined in the non-coincidence
gauge, we present the field equations in both the Jordan frame and the
Einstein frame.

To explore the effects of the conformal transformation on the physical
properties of solution trajectories within the conformal equivalent theories,
Section \ref{sec4} is dedicated to deriving precise solutions for the field
equations. We conduct a comparative analysis of the physical properties
between the two frames. It is observed that singular scaling solutions in one
frame correspond to singular scaling solutions in the other frame, displaying
identical asymptotic behaviour. Additionally, for the non-singular de Sitter
solution, it is established that the asymptotic behaviour of physical
properties remains unchanged under the application of the conformal
transformation. Moreover, in Section \ref{sec5}, we introduce an analytical
solution for the scalar-tensor theory in non-metricity gravity for the first
time. The analysis reveals that this universe originates from a Big Rip
singularity, transitions into an era characterized by an ideal gas, and
ultimately converges towards a de Sitter universe as a future attractor.
Notably, the observed behaviour of the physical parameters remains consistent
regardless of the frame in which the theory is defined. Finally, our findings
are summarized in Section \ref{sec6}.

\section{Symmetric teleparallel gravity}

\label{sec2}

Let $M^{n}$ be a manifold defined by the metric tensor, $g_{\mu\nu}$, and the
derivative $\nabla_{\lambda}$, defined by the generic connection $\Gamma
_{~\mu\nu}^{\lambda}$ with conditions, the $\Gamma_{\mu\nu}^{\lambda}$ to
inherit the symmetries of the metric tensor $g_{\mu\nu}$; that is, if $X$ is a
Killing vector of $g_{\mu\nu}$, i.e. $\mathcal{L}_{X}g_{\mu\nu}$, then
$\mathcal{L}_{X}\Gamma_{\mu\nu}^{\lambda}=0$, in which $\mathcal{L}_{X}$ is
the Lie derivative with respect the vector field $X$. Furthermore, for the
connection $\Gamma_{~\mu\nu}^{\lambda}$ it holds that the Riemann tensor
$R_{\;\lambda\mu\nu}^{\kappa}$ and torsion tensor $\mathrm{T}_{\mu\nu
}^{\lambda}~$are always zero; that is,
\begin{align}
R_{\;\lambda\mu\nu}^{\kappa}  &  \equiv\frac{\partial\Gamma_{\;\lambda\nu
}^{\kappa}}{\partial x^{\mu}}-\frac{\partial\Gamma_{\;\lambda\mu}^{\kappa}%
}{\partial x^{\nu}}+\Gamma_{\;\lambda\nu}^{\sigma}\Gamma_{\;\mu\sigma}%
^{\kappa}-\Gamma_{\;\lambda\mu}^{\sigma}\Gamma_{\;\mu\sigma}^{\kappa}=0,\\
\mathrm{T}_{\mu\nu}^{\lambda}  &  \equiv\Gamma_{\;\mu\nu}^{\lambda}%
-\Gamma_{\;\nu\mu}^{\lambda}=0.
\end{align}

In symmetric teleparallel theory of gravity only the non-metricity tensor
survives, defined as \cite{Nester:1998mp}
\begin{equation}
Q_{\lambda\mu\nu}=\nabla_{\lambda}g_{\mu\nu},
\end{equation}
that is,
\begin{equation}
Q_{\lambda\mu\nu}=\frac{\partial g_{\mu\nu}}{\partial x^{\lambda}}%
-\Gamma_{\;\lambda\mu}^{\sigma}g_{\sigma\nu}-\Gamma_{\;\lambda\nu}^{\sigma
}g_{\mu\sigma}.
\end{equation}

We define the disformation tensor%
\begin{equation}
L_{~\mu\nu}^{\lambda}=\frac{1}{2}g^{\lambda\sigma}\left(  Q_{\mu\nu\sigma
}+Q_{\nu\mu\sigma}-Q_{\sigma\mu\nu}\right)
\end{equation}
and the non-metricity conjugate tensor \cite{Hohmann}%
\begin{equation}
P_{~\mu\nu}^{\lambda}=\frac{1}{4}\left(  -2L_{~~\mu\nu}^{\lambda}+Q^{\lambda
}g_{\mu\nu}-Q^{\prime\lambda}g_{\mu\nu}-\delta_{(\mu}^{\lambda}Q_{\nu
)}\right)  ,
\end{equation}
where now the non-metricity vectors $Q^{\lambda}$ and $Q^{\prime\lambda}$ are
defined as%
\begin{equation}
Q_{\lambda}=Q_{\lambda~~~\mu}^{~~~\mu},Q_{\lambda}^{\prime}=Q_{~~\lambda\mu
}^{\mu}~,
\end{equation}
and%
\[
P^{\lambda}=P_{~\mu\nu}^{\lambda}g^{\mu\nu}=\frac{\left(  n-2\right)  }%
{4}\left(  Q^{\lambda}-Q^{\prime\lambda}\right)  .
\]

The non-metricity scalar is defined as
\[
Q=Q_{\lambda\mu\nu}P^{\lambda\mu\nu}
\]
and the gravitational Action Integral in STGR is given by the following
expression \cite{Nester:1998mp}%
\begin{equation}
S_{STGR}=\int d^{4}x\sqrt{-g}Q.
\end{equation}

The non-metricity scalar, $Q$, and the Ricciscalar~$\mathring{R}$ for the
Levi-Civita connection $\mathring{\Gamma}_{~\mu\nu}^{\lambda}$of the metric
tensor $g_{\mu\nu}$ differ by a boundary term $B$, that is, \cite{sc1}%
\begin{equation}
B=\mathring{R}-Q,
\end{equation}
where%
\begin{equation}
B=-\mathring{\nabla}_{\lambda}\left(  Q^{\lambda}-Q^{\prime\lambda}\right)
\end{equation}
and $\mathring{\nabla}_{\lambda}$ denotes covariant derivative with respect to
the Levi-Civita connection, $\mathring{\Gamma}_{~\mu\nu}^{\lambda}$.

\subsection{$f\left(  Q\right)  $-theory}

An extension of STGR which has drawn the attention recently is the $f\left(
Q\right)  $-gravity. In this theory, the gravitational Lagrangian is a
nonlinear function $f\left(  Q\right)  $, such that the Action Integral is
\cite{Koivisto2,Koivisto3}%
\[
S_{f\left(  Q\right)  }=\int d^{4}x\sqrt{-g}f\left(  Q\right)  \text{.}%
\]

The resulting gravitational field equations are%
\begin{equation}
f^{\prime}(Q)G_{\mu\nu}+\frac{1}{2}g_{\mu\nu}\left(  f^{\prime}%
(Q)Q-f(Q)\right)  +2f^{\prime\prime}(Q)\left(  \nabla_{\lambda}Q\right)
P_{\;\mu\nu}^{\lambda}=0, \label{feq1}%
\end{equation}
where $G_{\mu\nu}~$is the Einstein-tensor.

Moreover, connection $\Gamma_{~\mu\nu}^{\lambda}$ satisfies the equation of
motion%
\begin{equation}
\nabla_{\mu}\nabla_{\nu}\left(  \sqrt{-g}f^{\prime}%
(Q)P_{\phantom{\mu\nu}\sigma}^{\mu\nu}\right)  =0. \label{feq2}%
\end{equation}

When equation (\ref{feq2}) is satisfied for a given connection, we designate
the connection as defined in the coincidence gauge. Conversely, if the
equation is not satisfied, the connection is said to be defined in the
non-coincidence gauge. Furthermore, in the limit at which $f\left(  Q\right)
$ becomes linear, the field equations are reduced to those of symmetric
teleparallel gravity (STGR).

Last but not least, in the presence of a matter source minimally coupled to
gravity, the field equations (\ref{feq1}) are modified as follows%
\begin{equation}
f^{\prime}(Q)G_{\mu\nu}+\frac{1}{2}g_{\mu\nu}\left(  f^{\prime}%
(Q)Q-f(Q)\right)  +2f^{\prime\prime}(Q)\left(  \nabla_{\lambda}Q\right)
P_{\;\mu\nu}^{\lambda}=T_{\mu\nu},
\end{equation}
with the energy-momentum tensor $T_{\mu\nu}$ to give the degrees of freedom
for the matter source.

\subsection{Symmetric teleparallel scalar-tensor theory}

The symmetric teleparallel scalar-tensor theory is a Machian gravity, that is,
it satisfies Mach's principle, for which a scalar field non-minimally coupled
to gravity exists.

The gravitational Action Integral is \cite{sc1}%
\begin{equation}
S_{ST\varphi}=\int d^{4}x\sqrt{-g}\left(  \frac{F\left(  \varphi\right)  }%
{2}Q-\frac{\omega\left(  \varphi\right)  }{2}g^{\mu\nu}\varphi_{,\mu}%
\varphi_{,\nu}-V\left(  \varphi\right)  \right)  , \label{ai.01}%
\end{equation}
where $V\left(  \phi\right)  $ is the scalar field potential, which drives the
dynamics and $F\left(  \phi\right)  $ is the coupling function between the
scalar field and the gravitational scalar $Q$. The function, $\omega\left(
\phi\right)  $, can be eliminated with the introduction of the new scalar
field $d\Phi=\sqrt{\omega\left(  \varphi\right)  }d\varphi$. Hence, the Action
Integral (\ref{ai.01}) becomes%
\begin{equation}
S_{ST\Phi}=\int d^{4}x\sqrt{-g}\left(  \frac{F\left(  \Phi\right)  }{2}%
Q-\frac{1}{2}g^{\mu\nu}\Phi_{,\mu}\Phi_{,\nu}-V\left(  \Phi\right)  \right)  .
\end{equation}

The field equations which follow from the gravitational Action (\ref{ai.01})
are%
\begin{equation}
F\left(  \varphi\right)  G_{\mu\nu}+2F_{,\phi}\varphi_{,\lambda}P_{~~\mu\nu
}^{\lambda}+g_{\mu\nu}V\left(  \varphi\right)  +\frac{\omega\left(
\varphi\right)  }{2}\left(  g_{\mu\nu}g^{\lambda\kappa}\varphi_{,\lambda
}\varphi_{,\kappa}-\varphi_{,\mu}\varphi_{,\nu}\right)  =0,
\end{equation}%
\begin{equation}
\nabla_{\mu}\nabla_{\nu}\left(  \sqrt{-g}F\left(  \varphi\right)
P_{\phantom{\mu\nu}\sigma}^{\mu\nu}\right)  =0
\end{equation}
and%
\begin{equation}
\frac{\omega\left(  \varphi\right)  }{\sqrt{-g}}g^{\mu\nu}\partial_{\mu
}\left(  \sqrt{-g}\partial_{\nu}\varphi\right)  +\frac{\omega_{,\varphi}}%
{2}g^{\lambda\kappa}\varphi_{,\lambda}\varphi_{,\kappa}+\frac{1}{2}%
F_{,\varphi}Q-V_{,\varphi}=0.
\end{equation}

It is important to observe that for $\omega\left(  \varphi\right)  =0$,
$F\left(  \varphi\right)  =\varphi$, the latter field equations take the
functional form of $f\left(  Q\right)  $-theory \cite{sc1}, where now
$\varphi=f^{\prime}\left(  Q\right)  $ and $V\left(  \varphi\right)  =\left(
f^{\prime}(Q)Q-f(Q)\right)  $.

\section{Conformal transformation}

\label{sec2a}

The symmetric teleparallel scalar-tensor theory satisfies Mach's principle,
that is, the gravitational theory is defined in the Jordan frame. A similar
result holds for the $f\left(  Q\right)  $-theory. The Jordan frame is related
to the Einstein frame through a conformal transformation. This transformation
relates theories which are conformal equivalent. This equivalence it has to do
with the trajection solutions for the field equations, but it is not a
physical equivalence; since the physical properties of the theories do not
remain invariant under a conformal transformation. Conformal transformations
for the four-dimensional manifold were investigated in \cite{gg1}. Below we
consider a $n$-dimensional space.

Let $\bar{g}_{\mu\nu},~g_{\mu\nu}$ be two conformal equivalent metrics related
according to
\[
\bar{g}_{\mu\nu}=e^{2\Omega\left(  x^{\kappa}\right)  }g_{\mu\nu}~~,~\bar
{g}^{\mu\nu}=e^{-2\Omega\left(  x^{\kappa}\right)  }g^{\mu\nu}.
\]

Therefore, for the nonmetricity tensor we find%
\begin{equation}
\bar{Q}_{\lambda\mu\nu}=e^{2\Omega}Q_{\lambda\mu\nu}+2\Omega_{,\lambda}\bar
{g}_{\mu\nu}.
\end{equation}

Moreover,
\begin{equation}
\bar{Q}_{\mu}=\bar{Q}_{\mu~~\nu}^{~~\nu}=Q_{\mu}+2n\Omega_{,\mu},
\end{equation}%
\begin{equation}
\bar{Q}_{\mu}^{\prime}=\bar{Q}_{\mu\nu}^{~~~\nu}=Q_{\mu}^{\prime}%
+2\Omega_{,\mu}%
\end{equation}
and%
\[
\bar{P}^{\lambda}=\bar{P}_{~\mu\nu}^{\lambda}\bar{g}^{\mu\nu}=e^{-2\Omega
}P^{\lambda}+\frac{\left(  n-2\right)  \left(  n-1\right)  }{2}\Omega
^{,\lambda}.
\]

\bigskip

Therefore, for the non-metricity scalar we find%
\begin{equation}
\bar{Q}=\bar{Q}_{\lambda\mu\nu}\bar{P}^{\lambda\mu\nu}=e^{-2\Omega}Q+\left(
2\Omega_{,\lambda}P^{\lambda}+\left(  n-2\right)  \left(  n-1\right)
\Omega_{\lambda}\Omega^{,\lambda}\right)  .
\end{equation}

Consider now the Action Integral (\ref{ai.01}) for the n-dimensional
conformally related metric $\bar{g}_{\mu\nu}$, that is,%
\begin{equation}
\bar{S}_{ST\varphi}=\int d^{n}x\sqrt{-\bar{g}}\left(  \frac{F\left(
\varphi\right)  }{2}\bar{Q}-\frac{\omega\left(  \varphi\right)  }{2}\bar
{g}^{\mu\nu}\varphi_{,\mu}\varphi_{,\nu}-V\left(  \varphi\right)  \right)  .
\end{equation}

With respect to the metric \thinspace$g_{\mu\nu}$ and the conformal factor
$\Omega$, the latter Action Integral is%
\begin{align*}
\bar{S}_{ST\varphi}  &  =\int d^{n}x\sqrt{-g}\left(  e^{\left(  n-2\right)
\Omega}F\left(  \varphi\right)  \left(  \frac{Q}{2}+\Omega_{,\lambda
}P^{\lambda}\right)  \right) \\
&  +d^{n}x\sqrt{-g}\left(  e^{n\Omega}\left(  \frac{\left(  n-2\right)
\left(  n-1\right)  }{2}\Omega_{\lambda}\Omega^{,\lambda}-\frac{\omega\left(
\varphi\right)  }{2}e^{-2\Omega}g^{\mu\nu}\varphi_{,\mu}\varphi_{,\nu
}-V\left(  \varphi\right)  \right)  \right)  ,
\end{align*}
We select $F\left(  \varphi\right)  e^{\left(  n-2\right)  \Omega}=1$, that is
$\Omega=\frac{1}{2-n}\ln F\left(  \varphi\right)  $.

Therefore, the latter Action reads%
\begin{align}
\bar{S}_{ST\varphi}  &  =\int d^{n}x\sqrt{-g}\left(  \frac{Q}{2}%
+\Omega_{,\lambda}P^{\lambda}\right) \nonumber\\
~~  &  +\int d^{n}x\sqrt{-g}\left(  \left(  \frac{\left(  n-1\right)  F\left(
\varphi\right)  ^{\frac{n}{2-n}}}{2\left(  n-2\right)  F\left(  \varphi
\right)  }-\frac{\omega\left(  \varphi\right)  }{2}\frac{e^{\left(
n-2\right)  \Omega}}{F\left(  \varphi\right)  ^{\frac{n}{2-n}}}\right)
g^{\mu\nu}\varphi_{,\mu}\varphi_{,\nu}-F\left(  \varphi\right)  ^{\frac
{n}{2-n}}V\left(  \varphi\right)  \right)  .
\end{align}

The second terms become%
\begin{align*}
\int d^{n}x\sqrt{-g}\left(  \Omega_{,\lambda}P^{\lambda}\right)   &  =\int
d^{n}x\sqrt{-g}\left(  -\Omega\mathring{\nabla}_{\lambda}P^{\lambda}\right) \\
&  =\int d^{n}x\sqrt{-g}\left(  \frac{\left(  n-2\right)  }{4}\Omega B\right)
.
\end{align*}

We end with the gravitational Lagrangian%
\begin{equation}
\bar{S}_{ST\varphi}=\int d^{n}x\sqrt{-g}\left(  \frac{Q}{2}-\ln F\left(
\varphi\right)  \frac{B}{4}+\frac{A\left(  \varphi\right)  }{2}g^{\mu\nu
}\varphi_{,\mu}\varphi_{,\nu}-V\left(  \varphi\right)  F\left(  \varphi
\right)  ^{\frac{n}{2-n}}\right)
\end{equation}
with
\begin{equation}
A\left(  \varphi\right)  =\left(  \frac{\left(  n-1\right)  \left(
F_{,\varphi}\right)  ^{2}}{\left(  n-2\right)  F\left(  \varphi\right)
}-\frac{\omega\left(  \varphi\right)  }{F\left(  \varphi\right)  }\right)  .
\end{equation}

\section{Brans-Dicke Cosmology in Symmetric teleparallel theory}

\label{sec3}

Similarly to the consideration of the Brans-Dicke field in the
scalar-curvature theory, we take into account the following Action Integral
within a four-dimensional manifold in the context of symmetric teleparallel
theory. Indeed, in Action (\ref{ai.01}) we assume $F\left(  \varphi\right)
=\varphi$ and $\omega\left(  \varphi\right)  =\frac{\omega}{\varphi}%
,~\omega=const.$.

Thus, we arrive at the Lagrangian%

\begin{equation}
S_{BD\varphi}=\int d^{4}x\sqrt{-g}\left(  \frac{\varphi}{2}Q-\frac{\omega
}{2\varphi}g^{\mu\nu}\varphi_{,\mu}\varphi_{,\nu}-V\left(  \varphi\right)
\right)  .
\end{equation}
Parameter $\omega$ play a similar role as that of the Brans-Dicke parameter.

We define the new field $\varphi=e^{\phi}$, in order to write the latter
Action in the form of the Dilaton field
\begin{equation}
S_{D}=\int d^{4}x\sqrt{-g}e^{\phi}\left(  \frac{Q}{2}-\frac{\omega}{2}%
g^{\mu\nu}\phi_{,\mu}\phi_{,\nu}-\hat{V}\left(  \phi\right)  \right)
~,~\hat{V}\left(  \phi\right)  =V\left(  \phi\right)  e^{-\phi}. \label{sd.01}%
\end{equation}

On the other hand, in the Einstein frame, the equivalent Action integral is%
\begin{equation}
\bar{S}_{D}=\int d^{n}x\sqrt{-g}\left(  \frac{Q}{2}-\phi\frac{B}{4}+\frac
{\bar{\omega}}{2}g^{\mu\nu}\phi_{,\mu}\phi_{,\nu}-V\left(  \phi\right)
e^{-2\phi}\right)  ,~\bar{\omega}=\frac{3}{2}+\omega~,~\bar{V}\left(
\phi\right)  =V\left(  \phi\right)  e^{-2\phi}. \label{sd.02}%
\end{equation}

The solution trajectories of the field equations for the two gravitational
theories described by the Action integrals (\ref{sd.01}), (\ref{sd.02}) are
linked by the conformal transformation. However, no definitive conclusion can
be drawn concerning the relationship of the physical properties of the
solutions under the application of the conformal transformation.

The objective of this study is to examine how the conformal transformation
impacts the physical properties of the trajectory solutions in symmetric
teleparallel theory. To conduct such an analysis, we consider the background
geometry which describes an isotropic and homogeneous spatially flat FLRW
universe, with the line element
\begin{equation}
ds^{2}=-N^{2}\left(  t\right)  dt^{2}+a(t)^{2}\left(  dr^{2}+r^{2}\left(
d\theta^{2}+\sin^{2}\theta d\phi^{2}\right)  \right)  , \label{sd.03}%
\end{equation}
in which $a\left(  t\right)  $ is the scale factor and $N\left(  t\right)  $
is the lapse function. We derive the field equations for the two conformally
related models, namely $S_{D}$ and $\bar{S}_{D}$.

We obtain exact and analytic solutions for one of the models and thoroughly
examine the physical properties of these solutions. Subsequently, we apply the
conformal transformation to ascertain the corresponding exact and analytic
solutions for the second model, delineating the specific physical properties
of these solutions. Finally, we conduct a comparative analysis of the physical
properties between the solutions of the two conformally related theories.

For the spatially flat FLRW geometry described by the line element
(\ref{sd.03}) there are three families of symmetric connections which describe
a flat geometry and inherit the symmetries of the background space
\cite{ndim}. One family is defined in the coincidence gauge, for this family
the non-metricity scalar $Q$ has the same factional form with the torsion
scalar of teleparallelism. Thus, for the connection in the coincidence gauge
the symmetric teleparallel scalar-tensor theory is equivalent to the
scalar-torsion theory and $f\left(  Q\right)  $-theory is equivalent to
$f\left(  T\right)  $-theory.\ The remaining two families of connections are
defined in the non-coincidence gauge where, as it was found in \cite{ndim2}, a
scalar field is introduced into the gravitational theory which describes the connection.

In this piece of study we select to work in the framework of the connection
with nonzero components%
\begin{equation}
\Gamma_{\;tt}^{t}=\frac{\ddot{\psi}(t)}{\dot{\psi}(t)}+\dot{\psi}%
(t),\quad\Gamma_{\;tr}^{r}=\Gamma_{\;rt}^{r}=\Gamma_{\;t\theta}^{\theta
}=\Gamma_{\;\theta t}^{\theta}=\Gamma_{\;t\phi}^{\phi}=\Gamma_{\;\phi t}%
^{\phi}=\psi(t),
\end{equation}%
\begin{equation}
\Gamma_{\theta\theta}^{r}=-r~,~\Gamma_{\phi\phi}^{r}=-r\sin^{2}\theta
~,~\Gamma_{r\theta}^{\theta}=\Gamma_{\theta r}^{\theta}=\Gamma_{r\phi}^{\phi
}=\Gamma_{\phi r}^{\phi}=\frac{1}{r}~,
\end{equation}%
\begin{equation}
\Gamma_{\phi\phi}^{\theta}=-\sin\theta\cos\theta~,~\Gamma_{\theta\phi}^{\phi
}=\Gamma_{\phi\theta}^{\phi}=\cot\theta,
\end{equation}
in which $\dot{\psi}=\frac{d\psi}{dt}$, and without loss of generality we have
assumed that $N\left(  t\right)  =1$.

Thus, the non-metricity scalar is calculated
\begin{equation}
Q=-6H^{2}+9\dot{\psi}H+3\ddot{\psi}~,~\gamma=\dot{\psi}.
\end{equation}

We substitute into (\ref{sd.01}) and subsequently derive the cosmological
field equations in the Jordan frame, yielding:%

\begin{align}
3H^{2}+\frac{\omega}{2}\dot{\phi}^{2}+\frac{3}{2}\dot{\phi}\dot{\psi}%
-e^{-\phi}V\left(  \phi\right)   &  =0,\label{jo.01}\\
2\dot{H}+3H^{2}+2H\dot{\phi}-\frac{\omega}{2}\dot{\phi}^{2}-\frac{3}{2}%
\dot{\phi}\dot{\psi}-e^{-\phi}V\left(  \phi\right)   &  =0,\label{jo.02}\\
3\ddot{\psi}+2\omega\ddot{\phi}+H\left(  6\omega\dot{\phi}+9\dot{\psi}\right)
-6H^{2}+\omega\dot{\phi}^{2}-e^{-\phi}V_{,\phi}  &  =0,\label{jo.03}\\
\ddot{\phi}+\dot{\phi}^{2}+3H\dot{\phi}  &  =0, \label{jo.04}%
\end{align}
where $H=\frac{\dot{a}}{a}$ is the Hubble function.

Equations (\ref{jo.01})-(\ref{jo.04}) constitute a Hamiltonian dynamical
system described by the point-like Lagrangian,%
\begin{equation}
L\left(  a,\dot{a},\phi,\dot{\phi},\psi,\dot{\psi}\right)  =e^{\phi}\left(
3a\dot{a}^{2}+\frac{\omega}{2}a^{3}\dot{\phi}^{2}+\frac{3}{2}a^{3}\dot{\phi
}\dot{\psi}\right)  +a^{3}V\left(  \phi\right)  ,
\end{equation}
in which equation (\ref{jo.01}) is the constraint equation describing the
conservation law of \textquotedblleft energy\textquotedblright\ for the
classical Hamiltonian system. Recall that for $\omega=0$, the latter
Lagrangian reduces to that of $f\left(  Q\right)  $-gravity for the same connection.

We consider the conformally related metric,
\begin{equation}
d\bar{s}^{2}=-\bar{N}^{2}\left(  \tau\right)  d\tau^{2}+\alpha^{2}\left(
\tau\right)  \left(  dr^{2}+r^{2}\left(  d\theta^{2}+\sin^{2}\theta d\phi
^{2}\right)  \right)  ,
\end{equation}
with $a\left(  t\right)  =\alpha\left(  t\right)  e^{-\frac{\phi\left(
t\right)  }{2}}$, $N\left(  t\right)  =\bar{N}\left(  t\right)  e^{-\frac
{\phi\left(  t\right)  }{2}}$ and $d\tau=e^{-\frac{\phi\left(  t\right)  }{2}%
}dt$.

The field equations for the conformal equivalent theory (\ref{sd.02}) are%
\begin{align}
3\bar{H}^{2}-3\bar{H}\phi^{\prime}+\frac{\bar{\omega}}{2}\phi^{\prime2}%
+\frac{3}{2}\phi^{\prime}\psi^{\prime}-e^{-2\phi}V\left(  \phi\right)   &
=0,\label{ef.01}\\
2\bar{H}^{\prime}+3\bar{H}^{2}+3\bar{H}\phi^{\prime}-\frac{\bar{\omega}}%
{2}\phi^{\prime2}-\frac{3}{2}\phi^{\prime}\psi^{\prime}-e^{-2\phi}V\left(
\phi\right)   &  =0,\label{ef.02}\\
2\bar{H}^{\prime}+3H^{2}+\frac{2}{3}e^{-2\phi}\left(  V_{,\phi}-2V\right)
-\left(  \psi^{\prime\prime}+3\bar{H}\psi^{\prime}\right)   &
=0,\label{ef.03}\\
\phi^{\prime\prime}+3\bar{H}\phi^{\prime}  &  =0, \label{ef.04}%
\end{align}
where we have assumed $\bar{N}\left(  t\right)  =1$, and~$\alpha^{\prime
}=\frac{d\alpha}{d\tau},$ $\bar{H}\left(  \tau\right)  =\frac{\alpha^{\prime}%
}{\alpha}$ or $\bar{H}\left(  t\right)  =e^{-\frac{\phi}{2}}\left(
H+\frac{\dot{\phi}}{2}\right)  $.

Last but not least, the point-like Lagrangian for the field equations is
\begin{equation}
L\left(  \alpha,\alpha^{\prime},\phi,\phi^{\prime},\psi,\psi^{\prime}\right)
=3\alpha\alpha^{\prime2}-3\alpha^{2}\alpha^{\prime}\phi^{\prime}+\frac
{\bar{\omega}}{2}\alpha^{3}\phi^{\prime2}+\frac{3}{2}\alpha^{3}\phi^{\prime
}\psi^{\prime}+\alpha^{3}\bar{V}\left(  \phi\right)  .
\end{equation}

At this juncture, it is crucial to highlight that, for $\bar{\omega}=0$, the
latter gravitational theory is equivalent to the non-metricity theory with
boundary term, specifically with the $f\left(  Q,B\right)  =Q+f\left(
B\right)  $ theory of gravity \cite{fbo1}.

\section{Exact solutions}

\label{sec4}

In this Section we determine the existence of exact solutions for the field
equations that hold special significance. Additionally, we investigate the
physical properties of the solutions within the conformal equivalent theory.
Our focus is on determining the prerequisites for the existence of singular
solutions, corresponding to universes dominated by an ideal gas, as well as
identifying the conditions for a de Sitter solution. Subsequently, we utilize
the conformal transformation to deduce the exact solution in the second frame,
subsequently studying the physical properties and conducting a comparative
analysis of the solutions between the two frames.

\subsection{Singular solution in the Jordan frame}

We assume the scaling solution, $a\left(  t\right)  =a_{0}t^{p}$, $H\left(
t\right)  =\frac{p}{t}$, for the cosmological model defined in the Jordan
frame. This scale factor describes a universe dominated by an ideal gas with
equation of state parameter $w_{eff}=\frac{2-3p}{3p}$. Thus, for $p=\frac
{2}{3}$, the solution describes a universe dominated by a pressureless fluid,
i.e. dust. For $p=\frac{1}{2}$ it describes a universe dominated by radiation.
Moreover, for $p>1$ or $p<0$, the exact solution describes acceleration.

For the power-law singular solution $a\left(  t\right)  =a_{0}t^{p}$ and from
the equation of motion (\ref{jo.04}) it follows that
\begin{equation}
\phi\left(  t\right)  =\phi_{0}+\ln\left(  t^{1-3p}+\phi_{1}\right)
~,~p\neq\frac{1}{3} \label{pt.01}%
\end{equation}
and from the remaining of the field equations we derive%
\begin{equation}
V\left(  t\right)  =\frac{e^{\phi_{0}}p\left(  3p-1\right)  \phi_{1}}{t^{2}}
\label{pt.02}%
\end{equation}
and%
\begin{align}
\dot{\psi}  &  =\frac{1}{3\left(  3p-1\right)  }\left(  \frac{6p^{2}}%
{t}+2p\phi_{1}t^{3p-2}+\frac{\left(  1-3p\right)  ^{2}\omega}{t\left(
1+t^{3p-1}\right)  \phi_{1}}\right)  ~,~\phi_{1}\neq0,\\
\dot{\psi}  &  =\frac{1}{3\left(  3p-1\right)  }\frac{\omega\left(
6p-1\right)  -3p^{2}\left(  2+3\omega\right)  }{t}~,~\phi_{1}=0\text{. }%
\end{align}

Hence, from (\ref{pt.01}) and (\ref{pt.02}), we can write the potential
function as follows%
\begin{equation}
V\left(  \phi\right)  =e^{\phi_{0}}p\left(  3p-1\right)  \phi_{1}\left(
e^{\phi-\phi_{0}}-\phi_{1}\right)  ^{\frac{2}{2p-1}}.
\end{equation}

For the stiff fluid solution, i.e. $p=\frac{1}{3}$, we calculate%
\begin{equation}
\phi\left(  t\right)  =\phi_{0}+2\left(  \ln\frac{\sqrt{1+2\phi_{1}t^{2}}}%
{t}\right)  ~,~p=\frac{1}{3}%
\end{equation}%
\begin{equation}
V\left(  t\right)  =\frac{4e^{\phi_{0}}\phi_{1}}{t^{2}}~,~\dot{\psi}\left(
t\right)  =\frac{3+2\omega+4t^{2}\left(  \phi_{1}^{2}t^{2}+2\phi_{1}\right)
}{3t\left(  1+2\phi_{1}t^{2}\right)  }.
\end{equation}
Therefore the scalar field potential is%
\begin{equation}
V\left(  \phi\right)  =4\phi_{1}\left(  e^{\phi}-2\phi_{1}e^{\phi_{0}}\right)
.
\end{equation}

\subsubsection{Einstein frame}

We proceed now with the derivation of the exact solution for the conformally
related model in the Einstein frame. Therefore, after the application of the
conformal transformation we derive%
\begin{equation}
\alpha\left(  t\right)  =e^{\frac{\phi_{0}}{2}}t^{p}\sqrt{t^{1-3p}+\phi_{1}%
},~\bar{H}\left(  t\right)  =e^{-\frac{\phi_{0}}{2}}\left(  \frac
{1-p+2p\phi_{1}t^{3p-1}}{2t^{3p}\left(  t^{1-3p}+\phi_{1}\right)  ^{\frac
{3}{2}}}\right)  ~,~p\neq\frac{1}{3}.
\end{equation}
The equation of state parameter for the effective fluid, $\bar{w}_{eff}\left(
t\right)  =-1-\frac{2}{3}\frac{\bar{H}^{\prime}}{\bar{H}^{2}}$, is determined%
\begin{equation}
\bar{w}_{eff}\left(  t\right)  =\frac{3\left(  1-p\right)  t^{2-6p}-4p\left(
\left(  9p-5\right)  \phi_{1}t^{-3p}+\left(  3p-2\right)  \phi_{1}^{2}\right)
}{\left(  t^{3p+1}\left(  1-p\right)  +2p\phi_{1}\right)  ^{2}}.
\end{equation}
In the special limit for which $\phi_{1}=0$ and $V\left(  \phi\right)  =0$,
the latter expression becomes $\bar{w}_{eff}=1$, and easily we can write the
scale factor in terms of the new parameter $\tau$ as $\alpha\left(
\tau\right)  \simeq\tau^{\frac{1}{3}}$. Therefore for $\phi_{1}=0$, any
scaling solution in the Jordan frame corresponds to the scaling solution which
describes a stiff fluid in the Einstein frame.

On the other hand, for $\phi_{1}\neq0$ and for large values of $t$, it follows
that $\bar{w}_{eff}\left(  t\right)  \simeq-1+\frac{2}{3p}$~for $p>\frac{1}%
{3}\,$. \ This means that, in the asymptotic limit, the solution in the
Einstein and in the Jordan frames has the same physical properties. We recall
that $\tau\left(  t\rightarrow\infty\right)  \rightarrow\infty$ for $\phi
_{1}>0$ and $p>\frac{1}{3}$. Hence, as far as we move from the singularity the
two frames describe the same physical universe. In the contrary, near to the
singularity, that is $t\rightarrow0$, $\bar{w}_{eff}\left(  t\right)  \simeq1$.

For $p=\frac{1}{3}$, the solution at the Einstein frame is
\begin{equation}
\alpha\left(  t\right)  =e^{\frac{\phi_{0}}{2}}\sqrt{t^{2}+2\phi_{1}}%
~,~\bar{H}\left(  t\right)  =2\phi_{1}e^{\frac{\phi_{0}}{2}}\frac{t^{2}%
}{\left(  1+2t^{2}\phi_{1}\right)  ^{\frac{3}{2}}}%
\end{equation}
and%
\begin{equation}
\bar{w}_{eff}\left(  t\right)  =-\frac{1}{3}-\frac{2\phi_{1}}{3t^{2}}.
\end{equation}

Hence%
\begin{equation}
\bar{H}\left(  \alpha\right)  =\frac{1}{3}\left(  \frac{1}{\alpha}%
-\frac{e^{\phi_{0}}}{9\alpha^{3}}\right)  ~,~\bar{w}_{eff}\left(
\alpha\right)  =1+\frac{12\alpha^{2}}{9a^{2}-e^{\phi_{0}}}.
\end{equation}

Thus, for large values of time, the asymptotic solution resembles that of a
stiff fluid, similar to the scenario in the Jordan frame.

\subsection{de Sitter universe in the Jordan frame}

Consider now the de Sitter universe with $a\left(  t\right)  =a_{0}e^{H_{0}t}%
$, $H\left(  t\right)  =H_{0}$. Then from the field equations in the Jordan
frame we derive%
\begin{equation}
e^{\phi\left(  t\right)  }=e^{\phi_{0}}\left(  1-e^{-3H_{0}\left(  t-\phi
_{1}\right)  }\right)  ~
\end{equation}
and%
\begin{equation}
V\left(  \phi\right)  =3e^{\phi_{0}}H_{0}^{2}~,~\dot{\psi}=\frac{H_{0}\left(
2e^{3H_{0}t}-e^{3H_{0}\phi_{1}}\left(  3\omega+2\right)  \right)  }{3\left(
e^{3H_{0}t}-e^{3H_{0}\phi_{1}}\right)  }.
\end{equation}
This means that the de Sitter solution exists for constant potential function
$V\left(  \phi\right)  $.

\subsubsection{Einstein frame}

Now we transform the solutions in the Einstein frame. Indeed, the scale factor
and the Hubble function becomes%
\begin{equation}
\alpha\left(  t\right)  =\sqrt{e^{\phi_{0}}\left(  e^{2H_{0}t}-e^{-H_{0}%
t}e^{3H_{0}\phi_{1}}\right)  },
\end{equation}%
\begin{equation}
\bar{H}\left(  t\right)  =H_{0}\frac{e^{-\frac{\phi_{0}}{2}}e^{\frac{3}%
{2}H_{0}t}\left(  2e^{3H_{0}t}+e^{3H_{0}\phi_{1}}\right)  }{4\left(
e^{3H_{0}t}-e^{3H_{0}\phi_{1}}\right)  ^{\frac{3}{2}}}%
\end{equation}
while the effective equation of state parameter reads%
\begin{equation}
\bar{w}_{eff}\left(  t\right)  =\frac{12e^{3H_{0}\left(  t+\phi_{1}\right)
}-4e^{6H_{0}t}+e^{6H_{0}\phi_{1}}}{2\left(  e^{3H_{0}t}+e^{3H_{0}\phi_{1}%
}\right)  ^{2}}.
\end{equation}

Hence for large values of $t\rightarrow\infty$, it follows that $\bar{w}%
_{eff}\left(  t\right)  \simeq-1$, while for small values of $t\rightarrow0$,
we determine $\bar{w}_{eff}\left(  t\right)  \simeq1-\frac{24}{\left(
2+e^{3H_{0}\phi_{1}}\right)  ^{2}}+\frac{8}{2+e^{3H_{0}\phi_{1}}},$where for
$e^{3H_{0}\phi_{1}}\rightarrow0$, the limit $\bar{w}_{eff}\left(  t\right)
\simeq-1$ follows, and for $e^{3H_{0}\phi_{1}}\rightarrow\infty$ we
derive~$\bar{w}_{eff}\left(  t\right)  \simeq1$.

\subsection{Singular solution in the Einstein frame}

Consider now the scaling solution $\alpha\left(  \tau\right)  =\alpha_{0}%
\tau^{q}$, then from equations (\ref{ef.01})-(\ref{ef.04}) we derive%
\begin{equation}
\phi\left(  \tau\right)  =\bar{\phi}_{0}+\frac{\bar{\phi}_{1}}{1-3q}%
\tau^{1-3q}~,~q\neq\frac{1}{3},
\end{equation}%
\begin{equation}
\psi^{\prime}=-\frac{1}{3}\bar{\phi}_{1}\bar{\omega}\tau^{3q}-\frac{2q}%
{3\bar{\phi}_{1}\tau^{2}}\left(  \tau^{3q}-3\bar{\phi}_{1}\tau\right)
\end{equation}
and%
\begin{equation}
V\left(  \tau\right)  =\frac{q\left(  3q-1\right)  }{\tau^{2}}\exp\left(
\bar{\phi}_{0}+\frac{\bar{\phi}_{1}}{1-3q}\tau^{1-3q}\right)  ,
\end{equation}
or equivalently%
\begin{equation}
V\left(  \phi\right)  =\left(  3q-1\right)  \left(  \frac{\bar{\phi}_{1}%
}{\left(  1-3q\right)  }\right)  ^{\frac{2}{1-3q}}\left(  \phi-\bar{\phi}%
_{0}\right)  ^{\frac{2}{3q-1}}.
\end{equation}

In the case of $q=\frac{1}{3}$, the exact solution%
\begin{equation}
\phi\left(  \tau\right)  =\bar{\phi}_{0}+\bar{\phi}_{1}\ln\left(  \tau\right)
\end{equation}
follows, that is,%
\begin{equation}
\psi^{\prime}=-\frac{2\left(  1-3\bar{\phi}_{1}\right)  +3\left(  \bar{\phi
}_{1}\right)  ^{2}\bar{\omega}}{9\bar{\phi}_{1}\tau}~,~V\left(  \tau\right)
=0\,\text{. }%
\end{equation}

\subsubsection{Jordan frame}

For $q\neq\frac{1}{3}$, the solution at the Jordan frame is%
\begin{align}
a\left(  \tau\right)   &  =e^{-\phi_{0}}\exp\left(  \frac{\bar{\phi}_{1}%
}{3q-1}\tau^{1-3q}\right)  \tau^{q}~,~\\
H\left(  \tau\right)   &  =e^{\frac{\phi_{0}}{2}}\exp\left(  \frac{\bar{\phi
}_{1}}{3q-1}\tau^{1-3q}\right)  \tau^{-1-3q}\left(  q\tau^{3q}-\bar{\phi}%
_{1}\tau\right)  ,\\
w_{eff}\left(  \tau\right)   &  =\frac{q\left(  2-3q\right)  \tau
^{6q}+qt^{1+3q}\bar{\phi}_{1}+2\tau^{2}\left(  \bar{\phi}_{1}\right)  ^{2}%
}{3\left(  qt^{3q}-t\bar{\phi}_{1}\right)  ^{2}}\text{. }%
\end{align}

We remark that $w_{eff}\left(  \tau\rightarrow0\right)  \simeq-\frac{2}{3}$
and $w_{eff}\left(  \tau\rightarrow\infty\right)  \simeq-1+\frac{2}{3q}$.
Hence, far from the singularity, the physical properties of the solution
remain unchanged under the influence of the conformal transformation.

The case $q=\frac{1}{3}$ was studied before. Thus we omit it.

\subsection{de Sitter universe in the Einstein frame}

For the exponential scale factor $\alpha\left(  \tau\right)  =\alpha
_{0}e^{\bar{H}_{0}\tau}$, from the field equations (\ref{ef.01})-(\ref{ef.04})
in the Einstein frame we determine the exact solution%
\begin{equation}
\phi\left(  \tau\right)  =\bar{\phi}_{0}-\frac{3}{\bar{H}_{0}}\bar{\phi}%
_{1}e^{-3\bar{H}_{0}\tau}~,~V\left(  \tau\right)  =3\bar{H}_{0}^{2}\exp\left(
2\bar{\phi}_{0}-\frac{2}{3}\bar{\phi}_{1}e^{-3\bar{H}_{0}\tau}\right)  ,
\end{equation}%
\begin{equation}
\psi^{\prime}=2\bar{H}_{0}-\frac{\bar{\omega}}{3}\bar{\phi}_{1}e^{-3\bar
{H}_{0}\tau}.
\end{equation}
Therefore, the scalar field potential is%
\begin{equation}
V\left(  \phi\right)  =3\bar{H}_{0}^{2}.
\end{equation}

\subsubsection{Jordan frame}

Finally, in the Jordan frame the latter solution is
\begin{align}
a\left(  \tau\right)   &  =\exp\left(  \bar{H}_{0}t-\bar{\phi}_{0}+\frac
{\bar{\phi}_{1}}{3\bar{H}_{0}}e^{-3\bar{H}_{0}t}\right)  ~,\\
H\left(  \tau\right)   &  =\exp\left(  -3\bar{H}_{0}t+\frac{\bar{\phi}_{0}}%
{2}-\frac{\bar{\phi}_{1}}{6\bar{H}_{0}}e^{-3\bar{H}_{0}t}\right)  \left(
\bar{H}_{0}e^{3\bar{H}_{0}t}-\bar{\phi}_{1}\right)  ~
\end{align}
and%
\begin{equation}
w_{eff}\left(  \tau\right)  =-\frac{3e^{6\bar{H}_{0}t}\bar{H}_{0}^{2}%
+e^{3\bar{H}_{0}t}\bar{H}_{0}\bar{\phi}_{1}+2\left(  \bar{\phi}_{1}\right)
^{2}}{3\left(  \bar{H}_{0}e^{3\bar{H}_{0}t}-\bar{\phi}_{1}\right)  ^{2}}.
\end{equation}

From these expressions we have the limits $w_{eff}\left(  \tau\rightarrow
0\right)  =-\frac{3\bar{H}_{0}^{2}+\bar{H}_{0}\bar{\phi}_{1}+\left(  \bar
{\phi}_{1}\right)  ^{2}}{3\left(  \bar{H}_{0}-\bar{\phi}_{1}\right)  }$ and
$w_{eff}\left(  \tau\rightarrow\infty\right)  =-1$. We conclude that the de
Sitter universe is the asymptotic solution in the two frames.

The above discussion highlights that the solutions exhibit identical physical
properties in both the Jordan and Einstein frames at the asymptotic limits.
This observation is significant and sets it apart from the scalar-curvature or
scalar-torsion theories of gravity, where such equivalence does not hold true.

\section{Analytic solution}

\label{sec5}

In the preceding Section, we explored the existence of exact solutions for the
field equations. The derived solutions exhibit fewer degrees of freedom
compared to the original dynamical system, rendering them special or
asymptotic solutions. Subsequently, we proceed to establish the analytic
solution for the field equations. Specifically, for the Brans-Dicke field with
the potential function $V\left(  \phi\right)  =V_{0}\exp\left(  \left(
\lambda-1\right)  \phi\right)  $, we derive the analytic solution for the
field equations (\ref{jo.01})-(\ref{jo.04}). The field equations form a
three-dimensional Hamiltonian system with six degrees of freedom, enabling the
application of the Hamilton-Jacobi method to simplify the field equations and
to construct the analytic solution.

We consider the point transformation%
\begin{equation}
\ln a=\frac{1}{6}u~,~\phi=\Phi-\frac{u}{\lambda},~\psi=\psi,
\end{equation}
in which the Lagrangian function of the field equations is%
\begin{align}
L\left(  N,u,\dot{u},\Phi,\dot{\Phi},\psi,\dot{\psi}\right)   &  =\frac
{\exp\left(  \frac{\lambda-2}{2\lambda}u+\Phi\right)  }{12\lambda^{2}N}\left(
\left(  \lambda^{2}-6\right)  \dot{u}^{2}-6\lambda^{2}\dot{\Phi}\left(
\omega\dot{\Phi}+3\dot{\psi}\right)  +6\lambda\left(  2\omega\dot{\Phi}%
+3\dot{\psi}\right)  \right) \nonumber\\
&  ~~-V_{0}N\exp\left(  \exp\left(  \frac{\lambda-2}{2\lambda}u+\left(
\lambda-1\right)  \Phi\right)  \right)  ~. \label{ln.01}%
\end{align}
We have considered the lapse function $N\left(  t\right)  $ to be a
non-constant function, we see below that this necessary in order to write the
closed-form solution of the field equations.

From Lagrangian function (\ref{ln.01}) we can define the momentum%
\begin{equation}
p_{u}=\frac{\partial L}{\partial\dot{u}}~,~p_{\Phi}=\frac{\partial L}%
{\partial\dot{\Phi}}~,~p_{\psi}=\frac{\partial L}{\partial\dot{\psi}},
\end{equation}
that is,%
\begin{align}
\dot{u}  &  =-\frac{3N}{\lambda}\exp\left(  \frac{\lambda-2}{2\lambda}%
u-\Phi\right)  \left(  \lambda p_{u}+p_{\Phi}\right)  ,\label{ln.02}\\
\dot{\Phi}  &  =-\frac{N}{3\lambda^{2}}\exp\left(  \frac{\lambda-2}{2\lambda
}u-\Phi\right)  \left(  9\left(  \lambda p_{u}+p_{\Phi}\right)  +\lambda
^{2}p_{\psi}\right)  ,\label{ln.03}\\
\dot{\psi}  &  =-\frac{N}{9}\exp\left(  \frac{\lambda-2}{2\lambda}%
u-\Phi\right)  \left(  3p_{\Phi}-2\omega p_{\psi}\right)  . \label{ln.04}%
\end{align}

Therefore, the Hamiltonian function $\mathcal{H}=p_{q}\frac{\partial
L}{\partial\dot{q}}-L$ can be written%
\begin{equation}
\mathcal{H}\equiv N\exp\left(  \frac{\lambda-2}{2\lambda}u-\Phi\right)
\left(  36V_{0}\lambda^{2}e^{\lambda\Phi}-27\left(  \lambda p_{u}+p_{\Phi
}\right)  ^{2}-6\lambda^{2}p_{\Phi}p_{\psi}+2\lambda^{2}\omega p_{\Phi}%
^{2}\right)  =0, \label{hm.01}%
\end{equation}
where $\mathcal{H}=0$, follows from the constraint equation (\ref{jo.01}).

Consequently, Hamilton's equations are
\begin{equation}
\dot{p}_{u}=0~,~\dot{p}_{\psi}=0 \label{hm.02}%
\end{equation}
and%
\begin{equation}
\dot{p}_{\Phi}=2V_{0}\lambda e^{\lambda\Phi}N\exp\left(  \frac{\lambda
-2}{2\lambda}u-\Phi\right)  ,
\end{equation}
from which we infer that $p_{u}$ and $p_{\psi}$ are constants, that is
$p_{u}=p_{u}^{0}$, and $p_{\psi}=p_{\psi}^{0}$.

Let $S=S\left(  u,\Phi,\psi\right)  $ be the Action, then from (\ref{hm.01})
we can write the Hamilton-Jacobi equation%
\begin{equation}
\left(  36V_{0}\lambda^{2}e^{\lambda\Phi}-27\left(  \lambda\frac{\partial
S}{\partial u}+\frac{\partial S}{\partial\Phi}\right)  ^{2}-6\lambda^{2}%
\frac{\partial S}{\partial\Phi}\frac{\partial S}{\partial\psi}+2\lambda
^{2}\omega\left(  \frac{\partial S}{\partial\Phi}\right)  ^{2}\right)  =0.
\end{equation}
Moreover, from (\ref{hm.02}) it follows that $S\left(  u,\Phi,\psi\right)
=p_{u}^{0}u+p_{\psi}^{0}\psi+\hat{S}\left(  \Phi\right)  \,$, that is,%
\begin{equation}
\left(  36V_{0}\lambda^{2}e^{\lambda\Phi}-27\left(  \lambda p_{u}^{0}+\hat
{S}_{,\Phi}\right)  ^{2}-6\lambda^{2}p_{\psi}^{0}\hat{S}_{,\Phi}+2\lambda
^{2}\omega\left(  \hat{S}_{,\Phi}\right)  ^{2}\right)  =0. \label{hm.03}%
\end{equation}

Therefore
\begin{equation}
p_{\Phi}\equiv\hat{S}_{,\Phi}=-\lambda\left(  p_{u}^{0}+\frac{\lambda}%
{9}p_{\psi}^{0}\right)  \pm\frac{\left\vert \lambda\right\vert }{9}%
\sqrt{108V_{0}e^{\lambda\phi}+p_{\psi}^{0}\left(  18\lambda p_{u}^{0}+p_{\psi
}^{0}\left(  \lambda^{2}+6\omega\right)  \right)  }.
\end{equation}

Using the above mentioned expression, we can derive the action $\hat{S}\left(
\Phi\right)  $. The field equations (\ref{ln.02})-(\ref{ln.04}) are reduced to
the following dynamical system%
\begin{align}
\frac{1}{N}\dot{u}  &  =-\frac{3}{\lambda}\exp\left(  \frac{\lambda
-2}{2\lambda}u-\Phi\right)  \left(  \lambda p_{u}^{0}+p_{\Phi}\right)  ,\\
\frac{1}{N}\dot{\Phi}  &  =-\frac{1}{3\lambda^{2}}\exp\left(  \frac{\lambda
-2}{2\lambda}u-\Phi\right)  \left(  9\left(  \lambda p_{u}^{0}+\hat{S}_{,\Phi
}\right)  +\lambda^{2}p_{\psi}^{0}\right)  ,\label{hm.04}\\
\frac{1}{N}\dot{\psi}  &  =-\frac{1}{9}\exp\left(  \frac{\lambda-2}{2\lambda
}u-\Phi\right)  \left(  3\hat{S}_{,\Phi}-2\omega p_{\psi}^{0}\right)  .
\end{align}

We consider the new independent variable to the scalar field $\Phi$, such that
$u=u\left(  \Phi\right)  $ and $\psi=\psi\left(  \Phi\right)  $. Thus, the
analytic solution is expressed in terms of the closed-form functions
\begin{equation}
u\left(  \Phi\right)  =u_{0}+\lambda\Phi+\frac{2\lambda\sqrt{p_{\psi}^{0}}%
}{\sqrt{18p_{u}^{0}\lambda+p_{\psi}^{0}\lambda^{2}+6p_{\psi}^{0}\omega}%
}\arctan h\left(  \sqrt{\frac{108V_{0}e^{\lambda\Phi}+p_{\psi}^{0}\left(
18\lambda p_{u}^{0}+p_{\psi}^{0}\left(  \lambda^{2}+6\omega\right)  \right)
}{\sqrt{p_{\psi}^{0}\left(  18p_{u}^{0}\lambda+18p_{u}^{0}\lambda+p_{\psi}%
^{0}\lambda^{2}+6p_{\psi}^{0}\omega\right)  }}}\right)
\end{equation}
and
\begin{equation}
\psi\left(  \Phi\right)  =\psi_{0}+\frac{\lambda}{9}\left(  \ln\left(
108V_{0}e^{\lambda\Phi}\right)  +\frac{9p_{u}^{0}\lambda+p_{\psi}^{0}%
\lambda^{2}+6p_{\psi}^{0}\omega}{p_{\psi0}\lambda^{2}}\left(  u\left(
\Phi\right)  -u_{0}-\lambda\Phi\right)  \right)  .
\end{equation}

The Hubble function and the equation of state parameter $w_{eff}$ are
expressed as%
\begin{equation}
H\left(  \Phi\right)  =\frac{\dot{\Phi}}{N}\left(  \frac{1}{a}\frac{da}{d\Phi
}\right)  ~,~w_{eff}\left(  \Phi\right)  =-1-\frac{2}{3H^{2}}\frac{\dot{\Phi}%
}{N}\frac{dH}{d\Phi}.
\end{equation}

\begin{figure}[ptb]
\centering\includegraphics[width=1\textwidth]{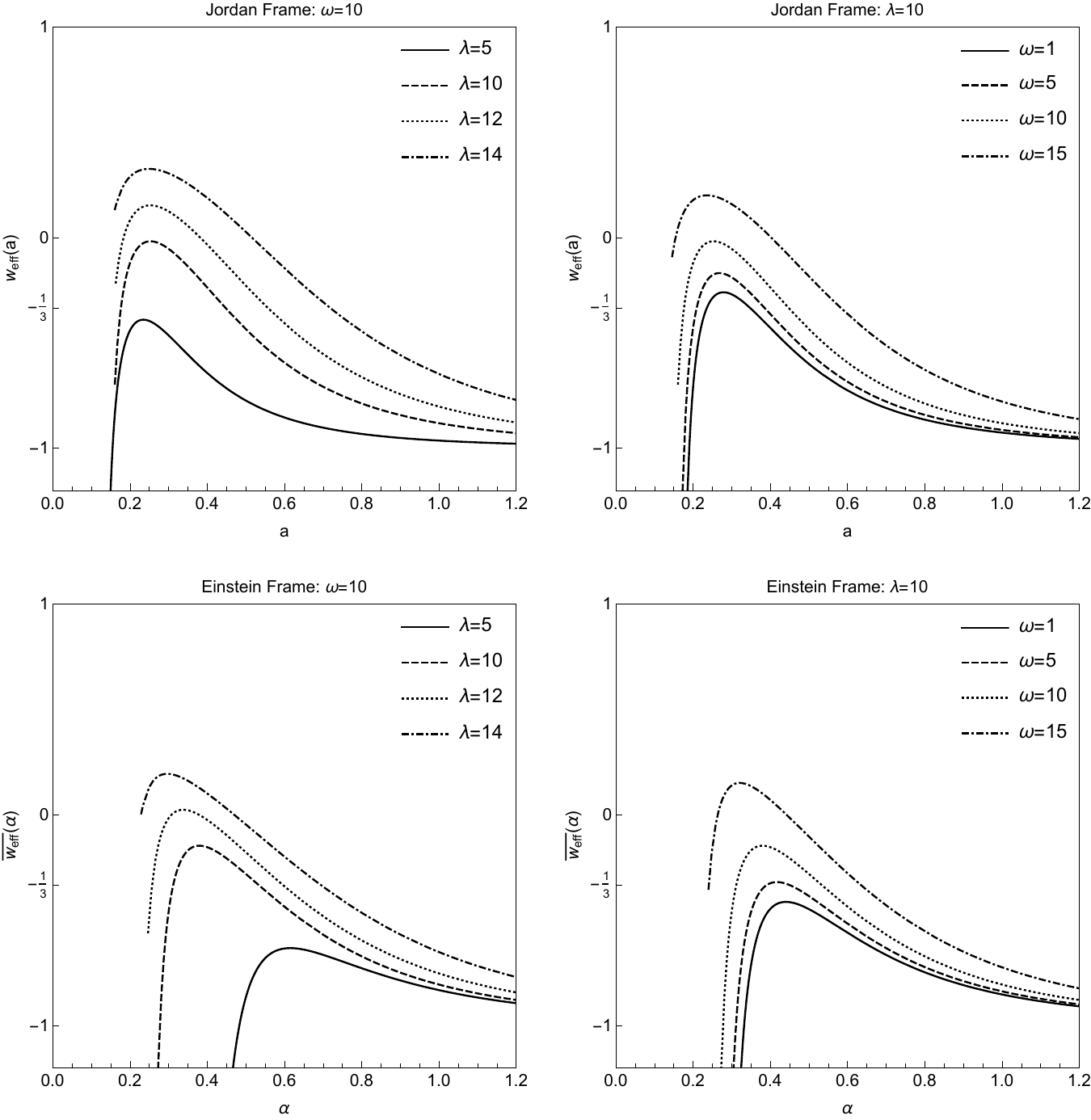}\caption{Qualitative
evolution of the effective equation of state parameter in the Jordan frame
$w_{eff}\left(  a\right)  $ and in the Einstein frame $\bar{w}_{eff}\left(
\alpha\right)  $ for different values of the free parameters. For all the
plots we consider the initial conditions$~\left(  p_{u}^{0},p_{\psi}^{0}%
,u_{0}\right)  =\left(  1,0.8,-10\right)  $, and $V_{0}=1$. We observe that
the behaviour for the equation of state parameter is similar in the two frames
and the de Sitter solution is a common future solution. }%
\label{fig1}%
\end{figure}

Figure \ref{fig1} illustrates the qualitative evolution of the equation of
state parameter, $w_{eff}\left(  \Phi\left(  a\right)  \right)  $, for the
above mentioned analytical solution, considering various values of the free
parameters. Additionally, we calculate and display the evolution of the
equation of state parameter $\bar{w}_{eff}\left(  \Phi\left(  \alpha\right)
\right)  $ for the conformal equivalent theory as defined in the Einstein
frame. The plots in both frames utilize identical values for the free
parameters, reflecting corresponding initial conditions.

It is observed that the universe initiates from a big rip singularity,
subsequently progresses towards a saddle point characterized by an ideal gas,
representing the matter-dominated era and finally transitions to the de Sitter
point. This behaviour is consistent across solution trajectories in both
frames, mirroring the findings for the asymptotic solutions in the preceding
section. While previously, the resemblance in the evolution of physical
parameters was noted at the asymptotic limits, Figure \ref{fig1} demonstrates
that this similarity persists throughout the global evolution of the
cosmological solution.

\section{Conclusions}

\label{sec6}

We performed an extensive analysis on the influence of the conformal
transformation on the physical properties of cosmological solution
trajectories within symmetric teleparallel gravity's conformal equivalent
theories. To undertake this analysis, we introduced the Brans-Dicke model in
the context of non-metricity gravity, alongside an analogue of the Brans-Dicke
parameter. Notably, when this parameter approaches zero, the non-metricity
scalar-tensor theory is reduced to the $f(Q)$-theory.

Regarding the background geometry, we focused on the isotropic and homogeneous
spatially flat FLRW metric. Concerning the theory's connection, we
specifically examined a connection defined within the non-coincidence gauge.
It is worth recalling that in the coincidence gauge, the cosmological field
equations simplify to those of scalar-torsion theory, limiting the new
information that could be deduced from this study. For this particular
cosmological model, we derived the field equations in both the Jordan and the
Einstein frames.

We derived exact solutions of particular significance in one frame,
illustrating both singular and non-singular solutions. Subsequently, we
utilized the conformal transformation to reconstruct the exact solutions for
the conformal equivalent theory. Our analysis involved a thorough comparison
of the physical properties for the two theories, each defined within different
frames. Notably, we discovered that the physical properties remained invariant
under the influence of the conformal transformation.

Consequently, singular solutions in one frame corresponded to singular
solutions in the other frame, displaying similar properties in the asymptotic
limit. Furthermore, we observed that the non-singular de Sitter solution
remained a de Sitter solution in the alternate frame as well. Furthermore, we
constructed for the fist time an analytic solution for the cosmological field
equations in non-metricity scalar-tensor theory. This solution describes an
cosmological model with Big Rip singularity, which involves to a matter
dominated solution and the final state of the universe is that of the de
Sitter universe. Surprisingly this specific cosmological history describes the
conformal equivalent theory. Hence, the physical equivalence of the physical
solutions between the two frames extends the asymptotic limits of the solutions.

In a future study we plan to investigate further such analysis by investigate
the case of compact objects.

\textbf{Data Availability Statements:} Data sharing is not applicable to this
article as no datasets were generated or analyzed during the current study.

\begin{acknowledgments}
The author thanks the support of Vicerrector\'{\i}a de Investigaci\'{o}n y
Desarrollo Tecnol\'{o}gico (Vridt) at Universidad Cat\'{o}lica del Norte
through N\'{u}cleo de Investigaci\'{o}n Geometr\'{\i}a Diferencial y
Aplicaciones, Resoluci\'{o}n Vridt No - 098/2022.
\end{acknowledgments}

\end{document}